Comments on the National Toxicology Program Report on Cancer, Rats and Cell Phone Radiation

Bernard J. Feldman

Department of Physics and Astronomy

University of Missouri-St. Louis, St. Louis, MO  63121

Abstract

With the National Toxicology Program issuing its final report on cancer, rats and cell phone radiation, one can draw the following conclusions from their data.  There is a roughly linear relationship between gliomas (brain cancers) and schwannomas (cancers of the nerve sheaths around the heart) with increased absorption of 900 MHz radiofrequency radiation for male rats.  The rate of these cancers in female rats is about one third the rate in male rats; the rate of gliomas in female humans is about two thirds the rate in male humans.  Both of these observations can be explained by a decrease in sensitivity to chemical carcinogenesis in both female rats and female humans.  The increase in male rat life spans with increased radiofrequency absorption is due to a reduction in kidney failure from a decrease in food intake.  No such similar increase in the life span of humans who use cell phones is expected.

This spring, the National Toxicology Program (NTP) issued its final report on their $25 million study on cancer in rats and mice exposed to 900 MHz cell phone radiation.  The study consisted of Sprague Dawley rats and B6C3F1 mice absorbing 900 MHz radiofrequency radiation at four different absorption levels:  0 W/kg, 1.5 W/kg, 3.0 W/kg and 6.0 W/kg.  The frequency of 900 MHz was chosen because it is typical for use in cell phones and other wireless devices. The exposure times were 10 minutes on and 10 minutes off for 18 hours a day, resulting in a total exposure of nine hours daily.  The animals were exposed whole body from in utero until two years of age.  The animals were monitored so that the exposure was at a low non-thermal or non-heating level.   Groups of 90 animals were used for each species, sex, modulation system, and absorption level.[1]

The study used two different radiofrequency modulation systems:  Global System for Mobile (GSM) and Code Division Multiple Access (CDMA).  The NTP results were reported for each modulation system separately.  However, I will argue below that the results from the two modulation systems should be combined.

A modulated electromagnetic wave as a function of time, t, can be expressed as

$E(t) = E_o \cos(\omega_o t)\cos(\omega_m t)$

$E_o$ is the amplitude of the wave, $\omega_o$ is the carrier frequency and $\omega_m$ is the modulation frequency. From trigonometry,

$$E(t) = 0.5 \, E_o \{\cos[(\omega_o + \omega_m)t] + \cos[(\omega_o - \omega_m)t]\}$$

Typically, $\omega_m$ is much smaller than $\omega_o$ and thus the modulation frequency can be neglected.

Looking at all the NTP data, there is no statistically significant difference between the GMS and the CDMA data, which provides strong experimental evidence for neglecting which modulation system was used.

    Combining the GMS and the CDMA data gives the results shown in Figures 1 and 2 for the number of gliomas and schwannomas detected in groups of 180 male and female rats as a function of the absorption of 900 MHz radiofrequency radiation. Gliomas are cancers of the brain and schwannomas are cancers of the nerve sheaths surrounding the heart. No cancers were observed in any of the mice. An explanation for all these results is given by this author in previous publications.[2]

    One can draw the following conclusions: there is roughly a linear relationship between cancer rates and radiofrequency absorption for schwannomas and gliomas in male rats. The cancer rate of female rats is about one third that of male rats (10 total cancers in female rats vs 31 total cancers in male rats).

    It is worth mentioning that the rate of gliomas (number of cancers per 100,000 of population) in human beings in the United States is about a third less in females than in males. This was true in 1980 before the wide spread usage of cell phones and was true in 2000 when cell phones were in wide use.[3] (The rate of gliomas roughly doubled from 1980 to 2000, but the interpretation of this observation is complicated by the advent of better detection systems.) The NTP study of bioassays observed that "male rats are more sensitive to chemical carcinogenesis compared to female rats."[4] May I suggest that this observation explains the difference in rates of gliomas and schwannomas in male and female rats and also the difference in the rates of gliomas in male and female humans. The connection between sensitivity to chemical carcinogenesis and exposure to radio frequency radiation is also spelled out in my previous publications.[2]

    The second interesting NTP result was that the irradiated male rats lived longer than the non-irradiated male rats and the greater the radiation absorption, the greater the life span.[1] The survival rate of male rats at 105 weeks increased from 28% at 0.0 W/kg to 62% at 6.0 W/kg. The effect in female rats was much smaller, from a survival rate at 105 weeks of 50% at 0.0 W/kg to 67% at 6.0 W/kg. The lead authors of the NTP study stated correctly that in the case of male rats, there was a trade-off between higher incidences of cancer and longer life spans. They then suggested that the same may be true of humans with their use of cell phones.[5] I disagree.

    The authors of the NTP study pointed out that the increase in male rat life spans was overwhelmingly due to a decrease in deaths from kidney failure—age-related nephropathy.[5] May I suggest the following explanation. The increased whole body absorption by male rats of the radiofrequency radiation leads to less food consumption needed to heat the body, which in turn leads to less by-products of food consumption, and thus less stress on the kidneys which have to remove those by-products from the blood. In the case of humans using cell phones, there is only localized heating of the head, much less total radiofrequency radiation absorbed, a far

smaller percentage reduction in food consumption and thus a negligible decrease in kidney stress. Also, unlike male rats, kidney failure is not a major factor in human life expectancies.

I hope these comments assist in the interpretation of the NTP results. Given that an increased rate of gliomas in humans due to cell phone usage has been reported by a number of workers,[6] the NTP study adds to the body of scientific data establishing and explaining this connection between radiofrequency radiation exposure and cancer in rats and humans.

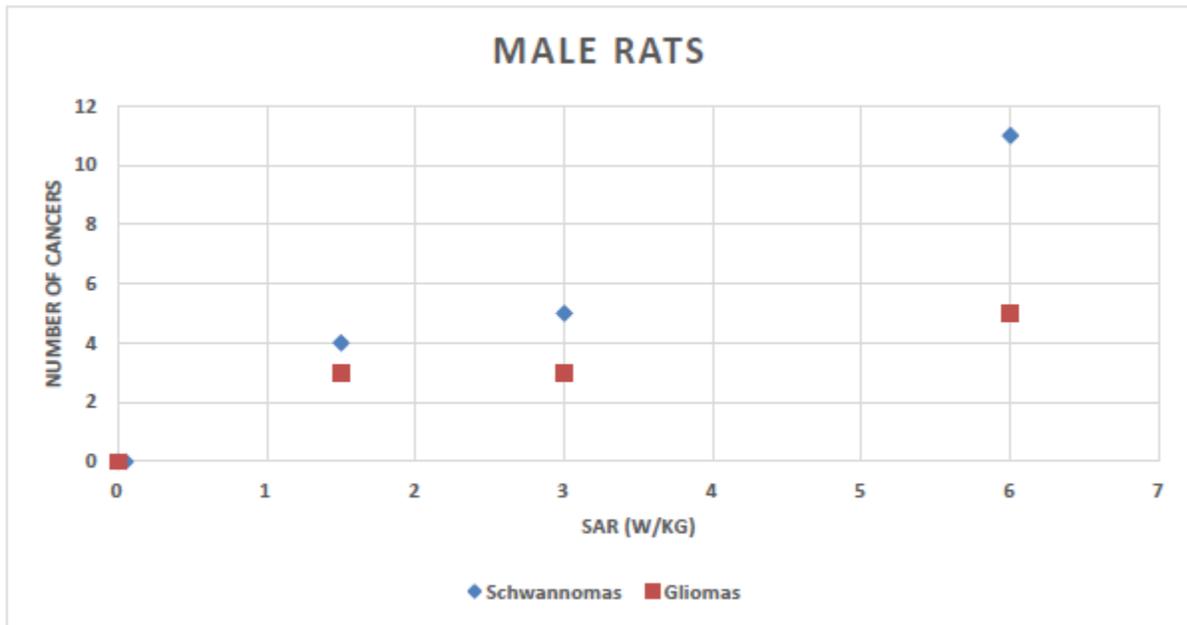

Figure 1: Number of cancers in 180 male rats as a function of the absorption of 900 MHz radio frequency radiation.

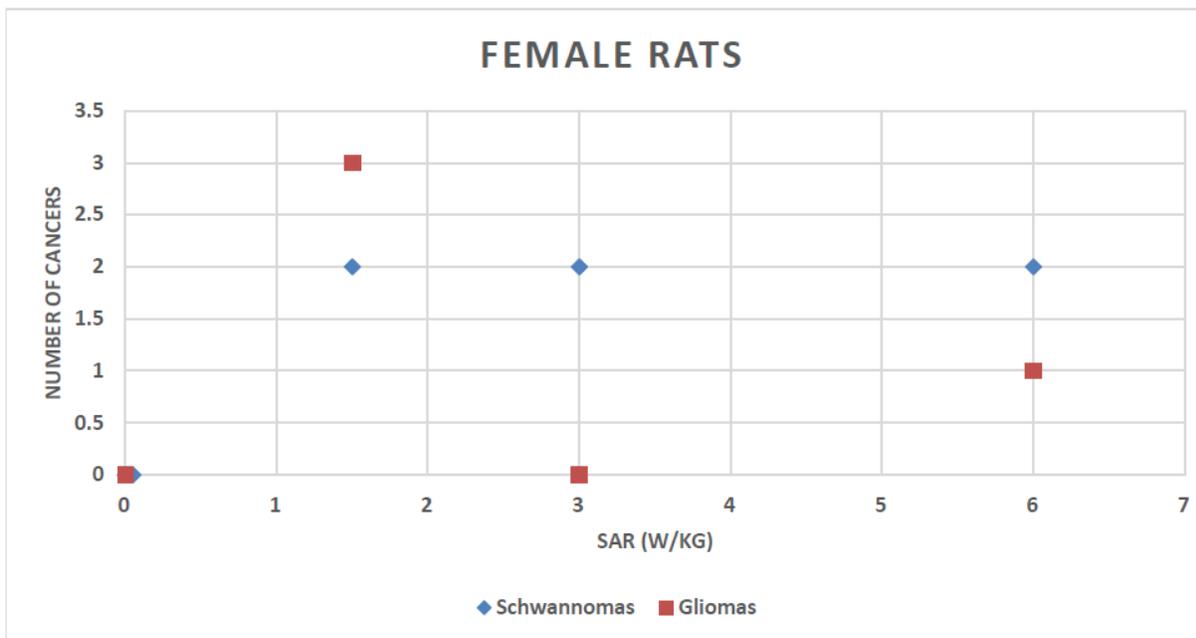

Figure 2: The number of cancers in 180 female rats as a function of the absorption of 900 MHz radio frequency radiation.